\documentclass[12pt,a4paper]{article}
\usepackage[latin1]{inputenc}
\usepackage{amsmath}
\usepackage{placeins}
\usepackage{natbib}
\usepackage{color,xcolor,setspace,geometry}
\setcitestyle{authoryear,open={(},close={)}}
\usepackage{cite}
\usepackage{amsfonts}
\usepackage{amssymb}
\usepackage{multirow}
\usepackage{graphicx}
\begin{document}
\begin{spacing}{1.35}

\title{The Hot Hand in Professional Darts}
	
\author{Marius \"Otting\thanks{Bielefeld University}, Roland Langrock$^*$, Christian Deutscher$^*$, \\ Vianey Leos-Barajas$^*$\thanks{Iowa State University}}

\date{}
	
\maketitle
\vspace{-2em}
    
\begin{abstract} We investigate the hot hand hypothesis in professional darts in a near-ideal setting with minimal to no interaction between players. Considering almost one year of tournament data, corresponding to 167,492 dart throws in total, we use state-space models to investigate serial dependence in throwing performance. 
In our models, a latent state process serves as a proxy for a player's underlying ability, and we use autoregressive processes to model how this process evolves over time. 
We find a strong but short-lived serial dependence in the latent state process, thus providing evidence for the existence of the hot hand. \end{abstract}

\section{Introduction}

In sports, the concept of the ``hot hand'' refers to the idea that athletes may enter a state in which they experience exceptional success. For example, in basketball, players are commonly referred to as being ``in the zone'' or ``on fire'' when they hit several shots in a row. However, in their seminal paper, \citet{gilovich1985hot} analyzed basketball free-throw data to find no support for a hot hand, hence coining the notion of the ``hot hand fallacy''. Since then, there has been mixed evidence, with some papers claiming to have found indications of a hot hand phenomenon and others disputing its existence.

There are mainly two types of approaches that have been used to investigate for potential hot hand patterns, namely 1) analyses of success rates conditional on the outcomes of previous attempts (see, e.g., \citealp{gilovich1985hot}, \citealp{dorsey2004bowlers}, \citealp{miller2014cold}), and 2) such that use a latent variable to describe the underlying ability (or ``hotness'') of a player (see, e.g., \citealp{sun2004detecting}, \citealp{wetzels2016}, \citealp{green2017hot}). Within 1), the hot hand is understood as a \textit{causal} relationship, where success increases the probability of success in subsequent attempts. In contrast, 2) focuses on \textit{correlation} in players' abilities, allowing for periods where players experience elevated success rates. In this paper, we focus on the latter, since this approach is more aligned with colloquial expressions such as ``being in the zone''. More specifically, using state-space models, we evaluate serial dependence in a latent state process, which can be interpreted as a player's varying ability.

Notably, \citet{miller2016surprised} highlight a subtle selection bias that may sneak into analyses of sequential data and challenge the findings of \citet{gilovich1985hot}.
Aside from mathematical fallacies, we note that many of the existing studies considered data, e.g.\ from baseball or basketball, which we believe are hardly suitable for analyzing streakiness in performances. For example, when analyzing hitting streaks of a batter in baseball, other factors such as the performance of the pitcher are also important but hard to account for. The same applies to basketball, as there are also several factors affecting the probability of a player to make a shot, e.g.\ the position (of a field goal attempt) or the effort of the defense. In particular, an adjustment of the defensive strategy to stronger focus on a player during a hot hand streak can conceal a possible hot hand phenomenon \citep{bocskocsky2014hot}.

To overcome these caveats, here we investigate whether there is a hot hand effect in professional darts, a setting with a high level of standardization of individual throws. In professional darts, well-trained players repeatedly throw at the dartboard from the exact same position and effectively without any interaction between competitors, making the course of play highly standardized. We consider a very large data set, with about $n=167,492$ throws in total, which allows for comprehensive inference on the existence and the magnitude of the hot hand effect. 

	\section{Modeling the Hot Hand in Darts}
    
    	\subsection{Data}
	
	Data was extracted from \texttt{http://live.dartsdata.com/}, covering all professional darts tournaments organized by the Professional Darts Corporation (PDC) between April 2017 and January 2018. In these tournaments, players start each leg with 501 points, and the first player to reach exactly zero points wins the leg. To win the match, a player must be the first to win a pre-specified number of legs (typically between 7 and 15). 
    In our analysis, we include all players who played at least 50 legs during the time period considered.
	
	At the beginning of legs, players consistently aim at high numbers to quickly reduce their points. The maximum score in a single throw is 60 as in a triple 20 (T20), but the data indicate the outcomes triple 19 (T19), triple 18 (T18), triple 17 (T17), triple 16 (T16), triple 15 (T15), and bullseye (Bull), to be targeted in the initial phase of a leg as well. Thus, in the initial phase of a leg we regard any throw to land in the set $H=\{\text{T15}, \text{T16}, \text{T17},\text{T18},\text{T19},\text{T20},\text{Bull}\}$ as success.
	A leg is won once a player reaches exactly 0 points, such that players do not target $H$ towards the end of legs, but rather numbers that make it easier for them to reduce to 0. To retain a high level of standardization and comparability across throws, we truncate our time series data, excluding throws where the remaining score was less than $c=180$ points. 
 
	We thus consider binary time series $\{ y_t^{p,l} \}_{t=1,\ldots,T_{p,l}}$, indicating the throwing success of player $p$ within his $l$--th leg in the data set, with 
	$$ y_t^{p,l} = 
	\begin{cases}
	1 & \text{if the $t$--th throw lands in $H$;} \\
	0 & \text{otherwise,}
	\end{cases}
	$$
	where the $T_{p,l}$--th throw is the last throw of player $p$ in his $l$--th leg with the player's remaining score still greater than or equal to $c=180$. The final data set comprises $n=167,492$ 
	throws of a dart, by $P=73$ players. To illustrate the structure as well as typical patterns of the data, we display Gary Anderson's throwing success histories throughout his first 15 legs in the data:
\begin{center} 
\begin{tabular}{l} 
001 011 011 \\
111 110 0 \\
000 111 101 \\
010 000 101 01 \\
000 110 101 \\
111 000 010 0 \\
110 100 101 \\
100 010 010 00 \\
101 010 000 1 \\
110 100 101 \\
101 101 1 \\
001 011 010 0 \\
000 010 010 11 \\
000 001 000 110 \\
000 111 100
\end{tabular} 
\end{center} 
Each row corresponds to one leg --- truncated when the score fell below 180 --- and gaps between blocks of three successive dart throws indicate a break in Anderson's play due to the opponent taking his turn. Next we formulate a model that enables us to potentially reveal any unusual streakiness in the data, i.e.\ a possible hot hand effect.
   
    \subsection{State-Space Model of the Hot Hand}
	
	We aim at explicitly incorporating any potential hot hand phenomenon into a statistical model for throwing success. Conceptually, a hot hand phenomenon naturally translates into a latent, serially correlated state process, which for any player considered measures his varying underlying ability. 
For average values of the state process, we would observe normal throwing success, whereas for high (low) values of the state process, we would observe unusually high (low) percentages of successful attempts. 
Figuratively speaking, the state process serves as a proxy for the player's ``hotness'' --- alternatively, it can simply be regarded as the player's varying ability. The magnitude of the serial correlation in the state process then indicates the strength of any potential hot hand effect. A similar approach was indeed used by \citet{wetzels2016} and by \citet{green2017hot}, who use discrete-state Markov models to measure the underlying ability. While there is some appeal in a discrete-state model formulation, most notably mathematical convenience and ease of interpretation (with cold vs.\ normal vs.\ hot states), 
we doubt that players traverse through only finitely many ability states, and advocate a continuously varying underlying ability state variable instead. 
Specifically, dropping the superscripts $p$ and $l$ for notational simplicity, we consider models of the following form:
    \begin{equation}
    y_t  \sim \text{Bern}\Bigl( \underbrace{\text{logit}^{-1}\bigl(\eta_t(s_t)\bigl)}_{=\pi_t}  \Bigr), 
	\quad s_t = h_t(s_{t-1}) + \epsilon_t, 
    \label{eq:ModellAllgemein}
    \end{equation}
	where $\{ y_t \}_{t=1,\ldots,T}$ is the observed binary sequence indicating throwing success, and $\{ s_t \}_{t=1,\ldots,T}$ is the unobserved continuous-valued state process indicating a player's varying ability. We thus model throwing success using a logistic regression model in which the predictor $\eta_t(s_t)$ for the success probability $\pi_t$ depends, among other things, on the current ability as measured by $s_t$. The unobserved ability process $\{s_t\}$ is modeled using an autoregressive process, and will include the possibility to be reduced to the nested special case of independent observations, corresponding to absence of any hot hand phenomenon. 
    
Model (\ref{eq:ModellAllgemein}) is a special case of a state-space model (SSM). Before we specify the exact forms of $\eta_t(s_t)$ and of $h_t$ in Chapter \ref{res}, in the next section we first discuss how to conduct maximum likelihood estimation within the general formulation given above.

    \subsection{Maximum Likelihood Estimation}
	The likelihood of a model as in (\ref{eq:ModellAllgemein}) involves analytically intractable integration over the possible realizations of $s_t$, $t=1,\ldots,T$. We use a combination of numerical integration and recursive computing, as first suggested by \citet{kit87}, to obtain an arbitrarily fine approximation of this multiple integral. Specifically, we finely discretize the state space, defining a range of possible values $[b_0, b_m]$ and splitting this range into $m$ intervals $B_i = (b_{i-1}, b_i)$, $i=1,\ldots,m$, of length $(b_m - b_0)/m$. 
   The likelihood of a single throwing history can then be approximated as follows:
	\begin{equation} 
	\begin{split}
	L_T & = \int \dots \int p(y_1, \dots, y_{_T}, s_1, \dots, s_{_T}) d_{s_T}\dots d_{s_1} \\
	& \approx \sum_{i_1=1}^{m} \dots \sum_{i_T=1}^{m} \Pr(s_1 \in B_{i_1}) \Pr(y_1 | s_1=b_{i_1}^\star) \\
    & \quad \times \prod_{t=2}^{T} \Pr (s_t \in B_{i_t} | s_{t-1}=b_{i_{t-1}}^\star) \Pr(y_t | s_t= b_{i_t}^\star),
	\end{split}
	\label{eq:likelihoodapprox}
	\end{equation}
	with $b_i^\star$ denoting the midpoint of $B_i$. This is just one of several possible ways in which the multiple integral can be approximated (see, e.g., \citealp{zucchini2016hidden}, Chapter 11).
In practice, we simply require that $m$ be sufficiently large. With the specification as logistic regression model as in (\ref{eq:ModellAllgemein}), we have that
 $$ \Pr(y_t | s_t= b_{i_t}^\star) = \left\{ \text{logit}^{-1}\bigl(\eta_t(b_{i_t}^\star)\bigl)\right\} ^{y_t} \cdot \left\{ 1 - \text{logit}^{-1}\bigl(\eta_t(b_{i_t}^\star)\bigl)\right\}  ^{1-y_t}.  $$
The approximate probability of the state process transitioning from interval $B_{i_{t-1}}$ to interval $B_{i_{t}}$, $\Pr (s_t \in B_{i_t} | s_{t-1}=b_{i_{t-1}}^\star)$, follows immediately from the specification of $h_t$ and the distribution of the noise, $\epsilon_t$.

The computational cost of evaluating the right hand side of Equation (\ref{eq:likelihoodapprox}) is of order $\mathcal{O}(Tm^T)$. However, the discretization of the state space effectively transforms the SSM into a hidden Markov model (HMM), with a large but finite number of states, such that we can apply the corresponding efficient machinery. 
In particular, for this approximating HMM, the forward algorithm can be applied to calculate its likelihood at a cost of order $\mathcal{O}(Tm^2)$ only \citep[Chapter 11]{zucchini2016hidden}. More specifically, defining ${\boldsymbol{\delta}} = (\delta_1,\ldots,\delta_m)$ with $\delta_i= \Pr (s_1 \in B_i)$, $i=1,\ldots,m$, the transition probability matrix (t.p.m) $\boldsymbol{\Gamma} = (\gamma_{ij})$ with $\gamma_{ij} = \Pr (s_t \in B_j |  s_{t-1}= b_i^\star)$, $i,j=1,\ldots,m$, and $m \times m$ diagonal matrix $\mathbf{P}(y_t)$ with $i$--th diagonal entry equal to $\Pr(y_t | s_t=b_i^\star)$, the right hand side of Equation (\ref{eq:likelihoodapprox}) can be calculated as 
	\begin{equation}\label{HMMlike}
	L_T \approx \boldsymbol{\delta} \mathbf{P}(y_1)\boldsymbol{\Gamma} \mathbf{P}(y_2)\dots\boldsymbol{\Gamma} \mathbf{P}(y_{T-1}) \boldsymbol{\Gamma} \mathbf{P}(y_{T}) \mathbf{1},
	\end{equation}
	with column vector $\mathbf{1}=(1,\ldots,1)' \in \mathbb{R}^m$. Equation (\ref{HMMlike}) applies to a single leg played by one player. Assuming independence of the individual leg's throwing histories, the complete likelihood, for the full data set, is obtained as the product of likelihoods of the form above:
	\begin{equation}
	L = \prod_{p=1}^{73} \prod_{l_p=1}^{L_p} \boldsymbol{\delta} \mathbf{P}(y_{1}^{p,l_p})\boldsymbol{\Gamma} \mathbf{P}(y_{2}^{p,l_p}) \dots\boldsymbol {\Gamma} \mathbf{P}(y_{T_{p,l}}^{p,l_p}) \mathbf{1}. 
	\label{eq:likelihood_fin}
	\end{equation}
	We estimate the model parameters by numerically maximizing the approximate likelihood, subject to the usual technical issues as detailed in \citet{zucchini2016hidden}.
	
	\section{Results}\label{res}
    Before presenting the results of the different hot hand models considered, we formulate two models that correspond to the hypothesis of no hot hand effect being present. These will serve as benchmarks for the SSMs to be considered below. 
    \textit{Model 1} assumes that each player's probability of success is constant across throws, i.e.\ the predictor in the logistic regression model for throwing success involves only player-specific intercepts:
    \begin{equation*}
    \text{logit} (\pi_t) = \beta_{0,p}.
    \label{mod:Benchmark1}
    \end{equation*}
Note we again suppress the superscripts $p$ and $l$ for player and leg, respectively, from $\pi_t$ for notational clarity. 
The estimated player-specific effects in \textit{Model 1}, $\beta_{0,1},\ldots,\beta_{0,73}$, range from $-0.857$ to $-0.135$, corresponding to throwing success probabilities ranging from $0.298$ to $0.466$.
    
    The relative frequency of hitting $H$, i.e.\ of throwing success in the early stages of a leg, does in fact differ notably across the three throws within a player's turn, with the empirical proportions of hitting $H$ in our data found to be $0.355$, $0.409$ and $0.420$ for the first, second and third throw, respectively. The substantial improvement after the first throw within a player's turn is due to the necessary re-calibration at the start of a turn. To take this into account, in \textit{Model 2} we include the categorical covariate $D_t$, $D_t \in \{1,2,3\}$, indicating the position of the dart thrown at time $t$ within the player's current turn (first, second or third):
    
%
    \begin{equation*}
    \text{logit} (\pi_t) = \beta_{0,p} + \beta_1 I_{\{ D_t=2 \} }  + \beta_2 I_{\{ D_t=3 \} },
    \label{mod:Benchmark2}
    \end{equation*}
with $I_{ \{ \cdot\}}$ denoting the indicator function, and $\beta_{0,p}$ player $p$'s baseline level for the first dart within any given turn.
For \textit{Model 2}, the coefficients $\beta_1$ and $\beta_2$, which correspond to the increase of throwing success probabilities after the first throw within a player's turn (on the logistic scale), are estimated as $0.228$ and $0.276$, respectively. The AIC clearly favors \textit{Model 2} over \textit{Model 1} ($\Delta \text{AIC} = 582$).

In \textit{Model 3}, we now include an underlying ability state variable $\{ s_t \}$, which we assume to follow an autoregressive process of order 1:
    \begin{equation*} 
	\begin{split}
	\text{logit} ( \pi_t)  &  = \beta_{0,p} + \beta_1 I_{\{ D_t=2 \} }  + \beta_2 I_{\{ D_t=3 \} } + s_t; \\ 
	\quad s_t & = \phi s_{t-1} + \sigma \epsilon_t, 
	\end{split}
    \label{eq:Modell3}
	\end{equation*}
with $\epsilon_t \stackrel{\text{iid}}{\sim} \mathcal{N}(0,1)$. Effectively this is a Bernoulli model for throwing success in which the success probability fluctuates around the players' baseline levels --- $\beta_{0,p}$, $\beta_{0,p} + \beta_1$ and $\beta_{0,p} + \beta_2$ for within-turn throws one, two and three, respectively --- according to the autoregressive process $\{ s_t\}$. The process $\{s_t\}$ can be interpreted as varying underlying ability (or ``hotness''). For $\phi = 0$, the model collapses to our benchmark \textit{Model 2} (i.e.\ absence of a hot hand), whereas $\phi>0$ would support the hot hand hypothesis. For the beginning of a leg, we assume $s_1 \sim \mathcal{N}(\mu_\delta,\sigma_\delta)$, i.e.\ that a player's underlying ability level starts afresh in every leg according to a normal distribution to be estimated. 
                
    
We fit \textit{Model 3} using $m=150$ and $-b_0 = b_m = 2.5$, monitoring the likely ranges of the process $\{ s_t \}$ to ensure the range considered is sufficiently wide given the parameter estimates. Table \ref{tab:resultsMod1} displays the parameter estimates (except the player-specific intercepts) including 95\% confidence intervals based on the observed Fisher information. 
Crucially, the estimate $\hat{\phi}=0.493$ supports the hot hand hypothesis, with the associated confidence interval not containing 0. This result corresponds to a considerable correlation in the underlying ability of the players' performances. The AIC clearly favors the hot hand model formulation, \textit{Model 3}, over the benchmark given by \textit{Model 2} ($\Delta \text{AIC} = 550$). 
However, the estimated mean of the initial distribution, $\hat{\mu}_\delta=-0.060$, indicates that players tend to start a leg with an ability level slightly below average.
This indicates that a momentum in performances may first of all need to be built, or in other words that the hot hand effect could be only short-lived, which is further discussed below.

	
    \begin{table}[ht]
\centering
\caption{Parameter estimates with 95\% confidence intervals for \textit{Model 3}.}\vspace{0.3em}
\label{tab:resultsMod1}
\begin{tabular}{cccc}
  \hline
 parameter & estimate & \multicolumn{1}{r}{95\% CI} \\ 
  \hline
$\phi$ & 0.493 & [0.437;\,0.549] \\ 
  $\sigma$ & 0.661 & [0.567;\,0.771] \\ 
  $\beta_1$ & 0.248 & [0.221;\,0.274] \\
  $\beta_2$ & 0.297 & [0.269;\,0.325] \\
  $\mu_\delta$ & -0.060 & [-0.010;\,-0.020] \\ 
  $\sigma_\delta$ & 0.700 & [0.658;\,0.745] \\ 
   \hline
\end{tabular}
\end{table}

To improve the realism of the hot hand model, we thus consider \textit{Model 4}, where we distinguish between transitions \textit{within} a player's turn to throw three darts (e.g.\ between first and second, or between second and third throw) and those \textit{across} the player's turns (e.g.\ between third and fourth throw). This extension accounts for the fact that there is a short break in a player's action between his turns, whereas within a single turn the three darts are thrown in very quick succession --- it thus seems plausible that any possible hot hand effect may show different time series dynamics within turns than across turns. 
\textit{Model 4} therefore assumes a periodic autoregressive process of order 1 (PAR(1); \citealp{franses2004periodic}):
    \begin{equation*}
	\begin{split}
	\text{logit} ( \pi_t)  &  = \beta_{0,p} + \beta_1 I_{\{ D_t=2 \} }  + \beta_2 I_{\{ D_t=3 \} } + s_t, \\ 
	\quad s_t & = \begin{cases}
\phi_a s_{t-1} + \sigma_a \epsilon_t & \text{if }\, t-1 \text{ mod } 3 = 0; \\
\phi_w s_{t-1} + \sigma_w \epsilon_t & \text{otherwise}.
\end{cases} 
	\end{split}
    \label{eq:Modell4}
	\end{equation*}
In the (approximate) likelihood, which still is of the form specified in (\ref{eq:likelihood_fin}), the t.p.m.\ $\boldsymbol{\Gamma}$ is then not constant across time anymore, but equal to either a within-turn t.p.m.\ $\boldsymbol{\Gamma}^{(w)}$ or  an across-turn t.p.m.\ $\boldsymbol{\Gamma}^{(a)}$.
For \textit{Model 4}, which is clearly favored over \textit{Model 3} by the AIC ($\Delta \text{AIC} = 242$), the parameter estimates as well as the associated confidence intervals are displayed in Table \ref{tab:resultsMod2}. The estimate of the persistence parameter of the AR(1) process active within a player's turn, $\hat{\phi}_w=0.726$, corresponds to quite strong correlation, which provides evidence for a clear hot hand pattern within turns. However, the estimate $\hat{\phi}_a = 0.057$ indicates only minimal persistence in the players' abilities across turns. In fact, when at time $t$ a player begins a new set of three darts within a leg, then the underlying ability variable is drawn from an $\mathcal{N}(0.057 s_{t-1},0.790^2)$ distribution, which is notably close to the initial distribution of the AR(1) process, an $\mathcal{N}(-0.034,0.690^2)$, which determines the underlying ability level at the start of a leg. In other words, there is a clear hot hand pattern, but the corresponding momentum is very short-lived and effectively only applies to darts thrown in quick succession. We cannot rule out that there may be a weak carry-over effect also across turns --- our results show no conclusive evidence in this regard. 

Table \ref{tab:allModels} provides an overview of the four models fitted, detailing the number of parameters, the AIC values, the type of state process (if any) and a short description. 
    
    
\begin{table}[ht]
\centering
\caption{Parameter estimates with 95\% confidence intervals for \textit{Model 4}. 
}\vspace{0.3em}
\label{tab:resultsMod2}
\begin{tabular}{rccc}
  \hline
 parameter & estimate & \multicolumn{1}{r}{95\% CI} \\ 
  \hline
$\phi_w$ & 0.726 & [0.642;\,0.811] \\ 
$\phi_a$ & 0.057 & [-0.010;\,0.125] \\ 
$\sigma_w$ & 0.464 & [0.353;\,0.609] \\
$\sigma_a$ & 0.790 & [0.700;\,0.893] \\
  $\beta_1$ & 0.270 & [0.242;\,0.297] \\
  $\beta_2$ & 0.330 & [0.301;\,0.359] \\
  $\mu_\delta$ & -0.034 & [-0.068;\,-0.001] \\ 
  $\sigma_\delta$ & 0.690 & [0.648;\,0.735] \\ 
   \hline
\end{tabular}
\end{table}

    \begin{table}[ht]
\centering
\caption{Overview of \textit{Models 1--4}.}\vspace{0.3em}
\label{tab:allModels}
\begin{tabular}{ccccp{6cm}}
  \hline
    & no.\ param. & AIC & state process & description \\ 
  \hline
\textit{Model 1} & 73 & 223,782 & -- & benchmark model containing only \newline player-specific intercepts \newline \\ \hline
\textit{Model 2} & 75  & 223,200 & -- & \textit{Model 1} + dummy variables for the throw within a players' turn \newline  \\ \hline
\textit{Model 3} & 79  &  222,650 & AR(1) & \textit{Model 2} + AR(1) state process for the underlying ability \newline \\ \hline
\textit{Model 4} & 81  & 222,398 & PAR(1) & \textit{Model 2} + PAR(1) state process, considering within and across transitions of a players' turn \\ 
   \hline
\end{tabular}
\end{table}

To obtain a more detailed picture of the short-term correlation found in the throwing performances, and also to check the goodness of fit of our models, in Table \ref{tab:resultsMod2cont} we compare the empirical relative frequencies of the eight possible throwing success sequences within players' turns --- 000, 001, 010, 011, 100, 101, 110, and 111 --- to the corresponding frequencies as expected under the four different models that were fitted. We restricted this comparison to the first two turns of players within each leg, and used Monte Carlo simulation to obtain the model-based frequencies of the eight sequences. 

\begin{table}[ht]
\centering
\caption{Relative frequencies of the eight possible throwing success histories within a player's turn. The second column gives the proportions found in the data, while columns 3--6 give the proportions as predicted under the various models fitted, for data structured exactly as the real data.}\vspace{0.3em}
\label{tab:resultsMod2cont}
\begin{tabular}{cccccc}
  \hline
 sequence & emp.\ prop.\ &  \textit{Model 1} & \textit{Model 2} & \textit{Model 3} & \textit{Model 4} \\ 
  \hline
0 0 0 & 0.252 & 0.225 & 0.222 & 0.239 & 0.250  \\ 
 0 0 1 & 0.151 & 0.144 & 0.159 & 0.153 & 0.150  \\ 
 0 1 0 & 0.130 & 0.144 & 0.152 & 0.139 & 0.136 \\ 
 0 1 1 & 0.114 & 0.094 & 0.111 & 0.113 & 0.110  \\ 
 1 0 0 & 0.103 & 0.144 & 0.121 & 0.114 & 0.109 \\ 
 1 0 1 & 0.080 & 0.094 & 0.088 & 0.082 & 0.082  \\ 
 1 1 0 & 0.086 & 0.094 & 0.084 & 0.083 & 0.082  \\ 
 1 1 1 & 0.084 & 0.062 & 0.063 & 0.075 & 0.083  \\ 
   \hline
\end{tabular}
\end{table}

The two benchmark models, \textit{Model 1} and \textit{Model 2}, which correspond to complete absence of any hot hand pattern, clearly underestimate the proportion of 000 and 111 sequences, with deviations of up to 0.030. This provides further evidence of correlation in throwing performances. 
\textit{Model 3} better reflects the cumulation of 000 and 111 sequences, with a maximum deviation of 0.013. Finally, \textit{Model 4}, which is favored by the AIC, almost perfectly captures the proportion of 000 and 111 sequences, with the main mismatch in proportions (0.006) found for 010 sequences. 

\begin{figure}[!htb]
\centering
\includegraphics[scale=0.8]{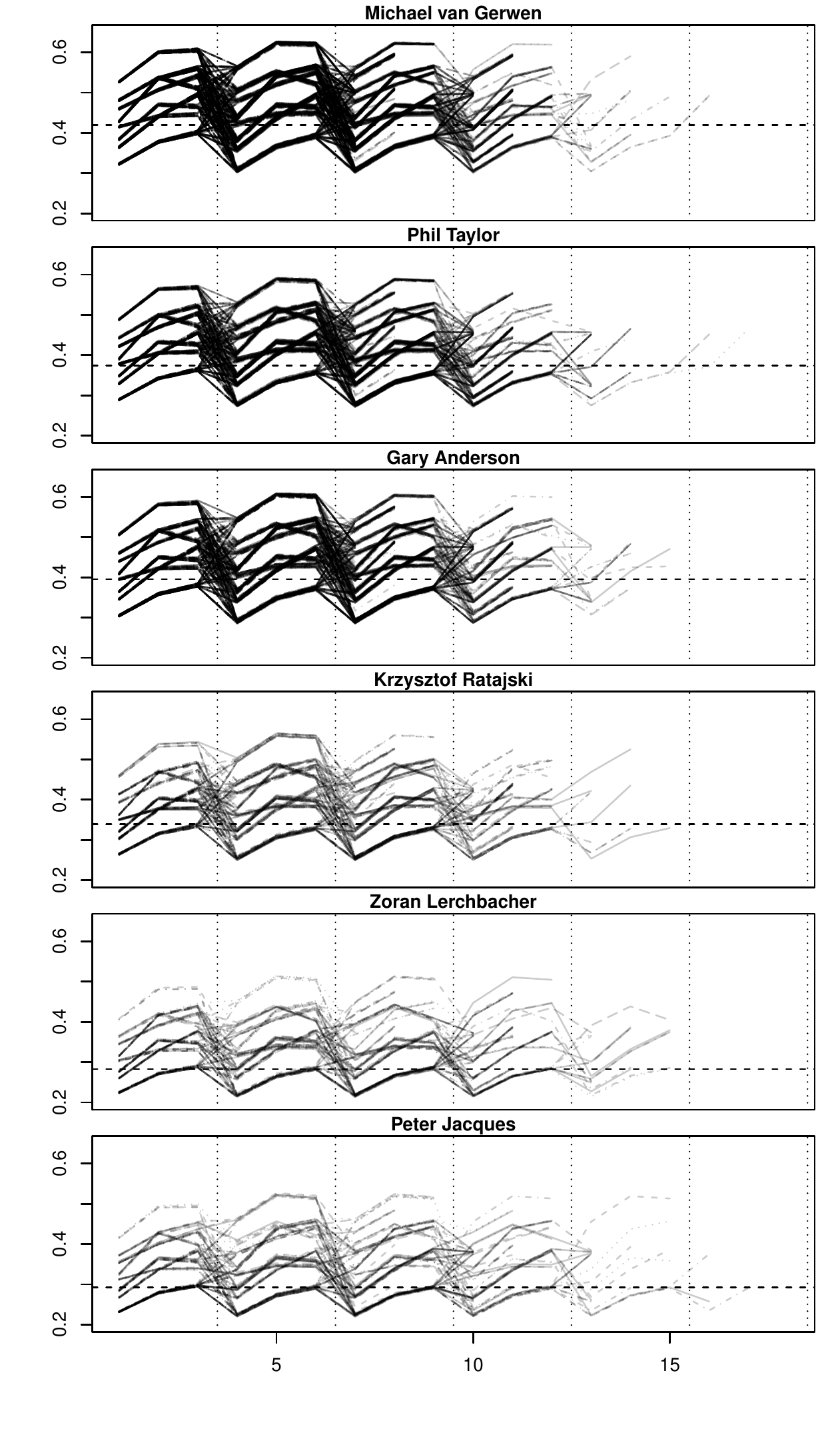}
\vspace{-2em}
\caption{Decoded most likely sequences of throwing success probabilities according to \textit{Model 4}, for $>100$ legs played by each of six players from the data set. The horizontal dashed lines indicate the player-specific intercepts for the respective player's within-turn throw one, and the vertical dashed lines denote the transition between a players' turn of three darts each.
} 
\label{fig:Viterbi}
\end{figure}

To further investigate typical patterns of the hidden process $\{s_t\}$ under \textit{Model 4}, we calculate the most likely trajectory of the latent (ability) state for each player and leg. Specifically, again dropping the superscripts $p$ and $l$, we seek 
$$ (s_1^*,\ldots,s_T^*) = \underset{s_1,\ldots,s_T}{\operatorname{argmax}} \; \Pr ( s_1,\ldots,s_T | y_1,\ldots, y_T ), $$
i.e.\ the most likely state sequence, given the observations. 
After discretizing the state space into $m$ intervals, maximizing this probability is equivalent to finding the optimal of $m^T$ possible state sequences. This can be achieved at computational cost $\mathcal{O}(Tm^2)$ using the Viterbi algorithm. 
We then calculate the corresponding trajectories ${\pi}_1^*,\ldots,{\pi}_T^*$ of the most likely success probabilities to have given rise to the observed throwing success histories, taking into account also the player-specific effects and the dummy variables. 
Figure \ref{fig:Viterbi} displays the decoded sequences for six players from the data set. 
Since there are only $2^3=8$ different possible sequences of observations within a player's turn, and since players start each turn almost unaffected by previous performances (cf.\ $\hat{\phi}_a = 0.057$), there is only limited variation in the \textit{most likely} sequences. 
The actual sequences may of course differ from these most likely sequences. The probability of hitting $H$ increases after the first throw within a turn due to the two dummy variables. We also see confirmed that the underlying ability level is not retained across turns.

	\section{Discussion}
Our analysis of a throwing success in darts provides strong evidence for a short-lived hot hand phenomenon. Our results indicate that within a player's turn, involving three darts thrown in quick succession, there is strong persistence in the underlying ability level. However, short breaks, which in the given setting result from the opponent taking his turn, effectively result in a fresh start of the process describing the player's underlying ability. These results provide new insights into the hot hand phenomenon, since previous studies did not explicitly account for possible breaks in a match. Explicitly accounting for such breaks in players' actions, such as in our \textit{Model 4}, can help to refine our understanding of the circumstances in which a hot hand can occur, including also the temporal scales involved. 
    
From a purely statistical point of view, if the hot hand phenomenon is understood as the presence of serial correlation in ability levels, then our findings provide strong evidence in favor of the hot hand. However, it is at least questionable whether serial correlation within a sequence of only three darts, thrown in quick succession, is what sports commentators, fans and athletes have in mind when referring to the hot hand. Instead, we believe that the notion of the hot hand is usually supposed to refer to players building up momentum over some period of a match. From that perspective, we would have expected to find (stronger) evidence of serial correlation also across players' turns. In other words, while we find strong serial correlation for within-hand throws, we do not actually find conclusive evidence for a hot hand the way it is usually understood.

Further research could focus on explicitly addressing player heterogeneity. In addition to the baseline level of $\pi_t$, the parameters $\phi_w$, $\phi_a$, $\sigma_w$ and $\sigma_a$, and hence the magnitude of the hot hand effect, may vary across players. This could reveal that for some players the hot hand effect lasts longer than for others, and potentially also across turns.
Modeling this individual variability could be achieved using covariates or, if no suitable covariates are available to explain the heterogeneity, via random effects. 

We also want to reiterate that the results presented in this paper refer to the hot hand as a correlational phenomenon, i.e.\ a correlation in the underlying ability level. Some previous studies have instead assumed the hot hand to be a causal phenomenon, where throwing success at time $t-1$ directly affects the probability of success at time $t$. With the binary time series data that we analyzed, we found that corresponding models that incorporate both correlational and causal effects could not be estimated reliably due to high numerical instability. With more detailed data on performances, we envisage approaches that allow both correlational and causal effects, in a single model, to potentially deliver important new insights into the hot hand concept.

\FloatBarrier

	\bibliography{refs}
    \bibliographystyle{plainnat}

\end{spacing}
\end{document}